\documentclass[11pt]{article}
\usepackage[utf8]{inputenc}
\usepackage{amsfonts,epsfig}
\usepackage[hyphens]{url}
\usepackage{breakurl}
\usepackage{comment}
\usepackage{color}
\usepackage{bbm}
\usepackage{microtype}
\usepackage{hyperref}
\hypersetup{
    colorlinks=true,
    linkcolor=black,
    citecolor=black,
    filecolor=black,
    urlcolor=black,
}

\usepackage{relsize}
\usepackage{fancyvrb}
\usepackage{amssymb,amsmath}
\usepackage{geometry}
\usepackage{sectsty}
\usepackage{caption}
\usepackage{subcaption}
\usepackage[normalem]{ulem}
\sectionfont{\sffamily\bfseries\upshape\large}
\subsectionfont{\sffamily\bfseries\upshape\normalsize}
\subsubsectionfont{\sffamily\mdseries\upshape\normalsize}
\makeatletter
\renewcommand\@seccntformat[1]{\csname the#1\endcsname.\quad}
\makeatother\renewcommand{\bibitem}{\vskip 2pt\par\hangindent\parindent\hskip-\parindent}

\makeatletter
\def\@maketitle{%
  \begin{center}%
  \let \footnote \thanks
    {\large \@title \par}%
    {\normalsize
      \begin{tabular}[t]{c}%
        \@author
      \end{tabular}\par}%
    {\small \@date}%
  \end{center}%
}
\makeatother

\title{\bf Limitations of ``Limitations of Bayesian leave-one-out cross-validation for model selection''\footnote{Discussion of Gronau and Wagenmakers (2018) for {\em Computational Brain \& Behavior}.  We thank Paul Bürkner for helpful comments, and the Academy of Finland, the Canadian Natural Sciences and Engineering Research Council, the  U.S. Office of Naval Research, National Science Foundation, and Defense Advanced Research Projects Administration for partial support of this work.}\vspace{.1in}}
\author{Aki Vehtari\footnote{Department of Computer Science, Aalto University, Finland}
\and Daniel P. Simpson\footnote{Department of Statisitcs, University of Toronto}
\and Yuling Yao\footnote{Department of Statistics, Columbia University.}
\and Andrew Gelman\footnote{Department of Statistics and Department of Political Science, Columbia University.} }
\date{12 Oct 2018\vspace{-.1in}}
\begin{document}\sloppy
\maketitle
\thispagestyle{empty}

\section{What was not but could be if}

The most important aspect of communicating statistical method to a new audience is to carefully and accurately sketch out the types of problems where it is applicable.  As people who think leave-one-out cross validation (LOO-CV or LOO for short) is a good method for model comparison and model criticism, we were pleased to discover that Gronau and Wagenmakers (2018, henceforth GW) chose to write a paper aimed at explaining the nuances of LOO methods to a  psychology audience. Unfortunately, we do not think the criticisms and discussions provided in their paper are so relevant to LOO as we understand it. The variant of LOO that GW discuss is at odds with a long literature on how to use LOO well; they focus on pathologizing a known and essentially unimportant property of the method; and they fail to discuss the most common issues that arise when using LOO for a real statistical analysis. In this discussion we try to discuss a number of concerns that everyone needs to think about before using LOO, reinterpret GW's examples, and try to explain the benefits of allowing for epistemological uncertainty when performing model selection.

\section{We need to abandon the idea that there is a device that will produce a single-number decision rule}

The most pernicious idea in statistics is the idea that we can produce a single-number summary of any data set and this will be enough to make a decision. This view is perpetuated by GW's paper, which says that the only way that LOO can provide evidence for choosing a single model is for the pseudo-Bayes Factor to grow without bound (or, equivalently, that the model weight approaches 1) as sample size increases. This is not a good way to use LOO and fundamentally misjudges both its potential and its limitations as a  tool for model selection and model criticism.

For a Bayesian model with $n$ data points $y_i \sim p(y | \theta)$ and parameters $\theta \sim p(\theta)$, LOO provides an estimate of the \emph{expected log posterior predictive distribution},
$$
\mathbb{E}_{\tilde{y}}\left(\log\left(\int\! p(\tilde{y}| \theta) p(\theta | y_\text{\rm all})\,d\theta\right)\right) \approx \frac{1}{n} \sum_{i=1}^n\log\left(\int\! p({y}_i| \theta) p(\theta | y_{-i})\,d\theta\right),
$$
where the expectation is taken with respect to new data, $y_\text{\rm all}$ is all n observed data points, and $y_{-i}$ is all data points \emph{except} the $i$th one. 

There are two things to note here. Firstly, the computed LOO score is an empirical approximation to the  expectation that we actually want to compute. This means that we must \emph{never} consider it as a single number without also considering its inherent  randomness. Considering both the LOO estimate and its empirical variance has been recommended since at least  Breiman et al.\ (1984, chapter 11). The model with higher estimated performance should not be automatically selected; rather, uncertainty in the estimates should be considered (see also Vehtari, Gelman, and Gabry, 2017).  

Since GW cite Vehtari, Gelman, and Gabry (2017) and Yao et al.\ (2018), we wish to emphasize that those papers of ours explicitly recommend computing the uncertainties due to not knowing the future data, and do  not recommend LOO weights as used in the experiments by GW.  Hence the claimed ``limitations of Bayesian leave-one-out cross-validation'' from GW do {\em not} apply to the version of Bayesian leave-one-out cross-validation that we actually recommend.  

At the very least, we recommend  that GW replace their pseudo-Bayes factors with  the pseudo-BMA+ weights of Yao et al.\ (2018), which assume a normal distribution of the empirical mean of the log predictive densities and makes the appropriate modification to the model weights.  Although this does not make a difference in their very specialized examples, we worry that people would see GW's work and assume that the method they describe is a ``best practice'' use of LOO rather than a version that happens to work in these extremely specialized situations. In Section \ref{sec:stack} we suggest  that stacking (Yao et al., 2018) is a much better way to define model weights as it uses the  LOO principle directly  rather than simply defining  ad hoc model weights.

The second thing to note is that the difference of LOO scores of two models can be decomposed into a sum of differences of log predictive densities for each left-out data point. This means that a two-way model comparison with LOO has more in common with a paired difference test than something like a Bayes factor.  Vehtari, Gelman, and Gabry (2017) argue that plots of these individual differences can be useful, especially as an exploratory tool, and can bring attention to cases where the preferred model is affected by  some unmodeled structure in the data.  Considering this decomposition does not lead to  a formal statistical test, but it can be useful when the aim of the analysis is to understand the data at hand rather than reach a ``discovery'' threshold.

This second property is extremely useful for applied work as it can be used to identify problems with model fit. As such, we recommend plotting the differences between the LOO log predictive densities even if LOO is not being used to perform model selection. Similarly, the $\hat{k}$ diagnostic produced by the \texttt{loo} package in R can be used as a generalized measure of leverage: if an observation has a $\hat{k}$ value larger than $0.7$ there is a very large difference between the predictive distribution that is conditioned on that observation and one that is not. These observations should be investigated as potentially problematic.  Gabry et al.\ (2018) has a detailed discussion of how plots generated in from LOO cross validation are a useful part of the applied Bayesian workflow.

Hence, we strongly feel that the question raised by GW is not ``Should we use LOO at all?'', but rather, ``Should we make model selection decisions based on LOO?''.

\section{Closed thinking in an $M$-open world:  Some tensions between reality and statistical convenience}

Cross-validation and LOO have many limitations. Gronau and Wagenmakers (2018) focus on a variation of one known limitation:  ``in the idealized case where there exists a data set of infinite size that is perfectly consistent with the simple model $M_S$, LOO will nevertheless fail to strongly endorse $M_S$.'' 

The strongest assumption in GW's paper is that the true model is one of the models under consideration, which Bernardo and Smith (1994) call the $M$-closed case. Cross-validation is usually recommended for the $M$-open case, where the true model is not included in the set of models compared.  In real life, we are just about always in the $M$-open scenario; like nearly all models, $M$-closed is a simplification of reality. While  it can sometimes be a useful simplification, we should never lose sight of the fact that methods that rely strongly on the $M$-closed assumption may be irretrievably biased.

Assuming that a model selection problem is $M$-closed allows the scientist to ignore epistemic uncertainty about their understanding of the process, and focus entirely on different models of aleatory uncertainty (the inherent randomness in the system). The extent to which this is a serviceable assumption will depend strongly on  context of.  With their exclusively focus on  $M$-closed model selection, GW have severely limited the scope of their paper compared to the real applicability of LOO methods.

When we estimate the predictive performance of a model for future data that are not yet available, we need to integrate over the true future data distribution $p_T(y)$, which is typically unknown. Vehtari and Ojanen (2012) provide an extensive survey of alternative scenarios including also different cases of conditional modeling $p_T(y|x)$, with stochastic, fixed or deterministic $x$. In the $M$-closed case, we know that one of the models is $p_T(y)$, but we don't know which one. In that case, it makes sense to model $p_T(y)$ as a posterior weighted average of all models under consideration, that is, replace $p_T(y)$ with a Bayesian model average (BMA) as proposed by San Martini and Spezzaferri (1984). This approach has the same model-selection-consistent behavior as the Bayes factor. Thus if the assumption is that one of the models is true, the BMA reference model approach  could be used, although asymptotic model selection consistency does not guarantee good finite sample performance, and BMA may have worse performance than LOO for small-to-moderate sample sizes even if one of the models is true.

On the other hand, in an $M$-open scenario, the BMA weights will still asymptotically converge to a single model (the one closest in Kullback-Leibler divergence to the data generating model), however this will no longer be the truth. In this case it is harder to make a case for using marginal likelihood-based weights.

Instead of using any explicit model or model averaging, LOO makes the assumption that future data can be drawn exchangeably from the same distribution as the observed data. This conditional independence structure is \emph{vital} to the success of LOO methods as it allows us to re-use observed-data as pseudo-Monte Carlo draws from the future data distribution (Bernardo and Smith, 1994). 

Cross-validation methods can have high variance when used on  data sets that don't fully represent the data distribution.  In such cases, some modeling assumptions about the future data distribution would be useful, in the same way that we use poststratification to adjust for extrapolations in sampling.

\section{If you use LOO better, you get more enlightening cases of failure} \label{sec:stack}

The main claims in GW is that when LOO is used for model selection (in a way that isn't in line with best practice) it is unable to select the true model and the LOO model weights are prior-dependent.  Their claims are true to some extent, but we don't feel they do a good job at representing the behaviour of LOO or are representative of how we'd recommend these tools are used.

The inconsistency of LOO model weights reported by GW for the first  two examples  is entirely an artifact of their poor application of the LOO principle. To wit: if you want to construct LOO model weights, they should be based on the weighted predictive distribution.   Yao et al.\ (2018) recommend using stacking weights, which choose the model weights that give the best expected log predictive density $$
\max_{\substack{w: w_k\geq 0, \\ \sum_{k=1}^K w_k=1}} \frac{1}{n}\sum_{i=1}^n \log\left(\sum_{k=1}^K w_k p(y_i| y_{-i},M_k)\right),
$$
where $M_k$ are the models under consideration. Yao et al.\ (2018) demonstrate with several $M$-open examples better performance of stacking compared to pseudo-BMA and BMA. 

Looking at the examples in GW, it turns out that stacking can also work well in the $M$-closed case.

In the beta-Bernoulli examples, the stacking weights can be derived analytically. The weights are $1$ for $H_0$ and $0$ for $H_1$. To see this,  we note that  model $H_0$ makes a perfect prediction, so $p(y_i | y_{-1}, H_0)=1$ and model $H_1$ also has a constant predictive performance $p(y_i | y_{-i}, H_1)= (a+n-1)/(a+n-1+b) =   c < 1$. The stacking objective function is  $n \log ( w_1 + w_2 c )$ which is maximized at $w_1= 1$ and $w_2=0$. This is independent of both the prior distribution and sample size $n$.
Moreover, when model $H_0$ does better than model $H_1$ in every data point (example 1) or every pair of data point consistently, the stacking weight corresponding to $H_0$ has to be $1$.

The lack of dependence on $n$ may look suspicious. But intuitively when each data point (example 1) or each pair of data point (example 2) uniformly supports $H_0$ more than $H_1$, it does not require $n \rightarrow\infty$ to conclude that $H_0$ dominates $H_1$. This conclusion instead follows from the cross-validation assumption that the current data is representative of the future data.

In examples 2 and 3 and idealized data, stacking with symmetric-leave-two-out would also converge faster due to the contrived nature of the data generation process. This is not to claim that stacking weights will generally give asymptotically consistent model selection in $M$-closed problems, just that it does for these two cases due to the simple data chosen by GW.

The examples that GW chose  are so specialized that the empirical mean of the log-predictive LOO densities has zero variance.  This is a  consequence of the data being both balanced and perfectly consistent with one of the hypotheses as well as the priors in examples 2 and 3 being centred on the mean of the data. These three properties imply that  $p(y_i | y_{-i})$ is constant for all $i$ and hence  Pseudo-BMA+ reduces to Pseudo-BMA used by GW. 

The zero variance of the log predictive densities and the lack of dependence on $n$ also demonstrates a clear problem with the cross-validation assumption that we can use subsets of existing data to approximate the distribution of future data.  The problem is that, for any finite $n$, the observed data sets in these examples are very strange (either all $1$s or all $\{0,1\}$ pairs) with zero variance. The LOO assumption will then assume that \emph{all} future data is also like this with zero variance, which is problematic.  Because even though we get consistency of the stacking weights in this case, if $H_0$ was $\theta =1-\epsilon$ for some small $\epsilon>0$, there would be a very high chance of observing a data set that was all $1$s even though the future data would include some values that are $0$.  This means that LOO will be biased when the true model lies very close to an edge case.

The experiments in GW could be modified to better illustrate the limitation of LOO by considering cases where the complex model is correct. Consider the following alternative experiments where the more complex model is true:
\begin{itemize}
\item In example 1, let the true $\theta_T= 1-\epsilon$, but for the simpler model keep $\theta_0=1$.
\item In example 2, let the true $\theta_T= \frac{1}{2}+\epsilon$, but for the simpler model keep $\theta_0=\frac{1}{2}$.
\item In example 3, let the true $\theta_T= \epsilon$, but for the simpler model keep $\theta_0=0$.
\end{itemize}
If we choose $\epsilon$ very small (but within the limits of the floating point accuracy for the experiments), we should see the same weights as in the original experiments as long as we observe the same data, and only when we occasionally observe one extra 0 in example 1, one extra 1 in example 2, or extra positive value in 3 we would see differences.

One way to illustrate the limitations of LOO (and cross-validation and information criteria in general), would be to compare the behaviour of the LOO stacking weights as  $\epsilon$ moves from very small to much larger, plotting how large $\epsilon$ needs to be before we see that the more complex model is strongly favored.

We know that we can get the standard deviation of the parameter posterior smaller than $\epsilon$ with $\mathcal{O}(\epsilon^{-2})$ data points. However because, we need enough data to be confident about the distribution of future data, we expect to require far more data points than that for the stacking weights to confidently select the correct model in the above $\epsilon$-experiment.  See the demonstration in Vehtari (2018a) and related results by Wang and Gelman (2014). 

This is the price we pay for not trusting any model and thus not getting benefits of reduced variance through the proper modeling of the future data distribution! This variability makes cross-validation bad for model selection when the differences between the models are small, and it just gets worse in case of a large number of models with similar true performance (see, e.g., Piironen and Vehtari, 2017). Even with just two models to compare, cross-validation has also a limitation that the simple variance estimate tend to be optimistic (Bengio and Grandvalet, 2004).

\section{It really helps to look at the estimates}
 GW demonstrate that LOO is not consistent for model selection by showing three examples where the pseudo-BMA weight for the true model does not approach one as more data arrives. We feel there is a more nuanced way to look at this: namely the models that GW compare are nested, which means that the more complex model contains the simpler model.  This means that the posterior predictive distribution from the more complex model will eventually be the same as the posterior predictive distribution constructed from the simpler model if the data is consistent with the simpler model.  One way to interpret  this phenomenon is that the two hypotheses are no longer distinct after we see enough data.   Viewed this way,  the prior dependence that GW demonstrated was just the prior choice slightly affecting the speed at which the two hypotheses are converging to each other. 
 
A quick look at the parameter estimates in any of the three GW models shows that in each case the posterior for $\theta$  is rapidly converging on the specific value of $\theta$ (respectively $1$, $1/2$, and $0$)  that defines the null model.  For concreteness, let's consider example 2, where the difference between the models is that the null model sets $\theta=1/2$ while the alternative model uses $\theta \sim \mathcal{B}(a,b)$.  We do not see a large difference between reporting ``$H_0$ best models the data'' or  ``$H_1$ best models the data and $\theta \approx 1/2$''.
 
Our recommendation is that if the LOO comparison taking into account the uncertainties says that there is no clear winner, then neither of the models should be selected and instead model averaging or expansion should be used. If two models are being compared and they give similar predictions, then it shouldn't matter which one is used. Unlike GW, we typically prefer to use the more complex one.

There is some tension between our preference for predictive models and GW's implied interest in discriminative ones.  GW's preference for parsimony appears to come in part from their desire to make statements like ``x is not associated with y.''  Whereas, when we use our models to make statements about future data, we would prefer the more complex model in the nested model in order to be certain that uncertainties are not underestimated.

We don't recommend using the more complex model thoughtlessly. We require strict model checking and calibration of the complex model, and then proceed using projective variable selection (Goutis and Robert, 1998; Dupuis and Robert, 2003; Piironen and Vehtari, 2017; Piironen et al., 2018) to decide if some parts of the model can be dropped safely. Overfitting can be viewed as the use of too flat a prior distribution---but we and others use flat, or relatively flat, priors all the time, so overfitting is a real and persistent concern.  Beyond concerns of stability in estimation and prediction, simpler models can also be motivated by a desire to 
reduce measurement costs in the future or to make it easier to explain the model for application experts, but this decision task then includes implicit or explicit costs for the measurements or complexity of the explanation.  This final step is not considered by GW, but we consider it to be a \emph{vital} part of any workflow that attempts model selection using LOO.

To clarify, we recommend using the encompassing model if (a) the model is being used to make predictions, (b) the encompassing model has passed model checking, and (c) the inference has passed diagnostic checks.  From a Bayesian perspective, a larger and smaller model can be considered as two different priors on a single larger set of distributions, and there is no reason why the larger, less restrictive model, should be associated with flat priors for the additional parameters, even though this has often been the tradition in model comparison.

Bayesian theory says that we should integrate over the uncertainties. The encompassing model includes the submodels, and if the encompassing model has passed model checking, then the correct thing is to include all the models and integrate over the uncertainties.  For the encompassing model to pass model checks, good priors will typically be required that appropriately penalize the complexity of the model (Simpson et al., 2018).  If the models have similar cross-validation performance, the encompassing model is likely to have thicker tails of the predictive distribution, meaning it more cautious about rare events. We think this is good.

Here are some reasons why it is common in practice to favor more parsimonious models:
\begin{itemize}
\item The maximum likelihood inference is common and it doesn't work well with more complex models. Favoring the simpler models is a kind of regularization.
\item Bad model misspecification. Bayesian inference can perform poorly if the model has major problems, and with complex models there are more possibilities for misspecifing the model and the misspecification even in one part can have strange effects in other parts.
\item Bad priors. Actually priors and models are inseparable, so this is kind of same as the previous one. It is more difficult to choose good priors for more complex models, because it's difficult for humans to think about high dimensional parameters and how they affect the predictions. Favoring the simpler models can avoid the need to think harder about priors.
\end{itemize}

We agree with Neal (1996), who wrote:
\begin{quotation}\noindent
Sometimes a simple model will outperform a more complex model [\dots] Nevertheless, I believe that deliberately limiting the complexity of the model is not fruitful when the problem is evidently complex. Instead, if a simple model is found that outperforms some particular complex model, the appropriate response is to define a different complex model that captures whatever aspect of the problem led to the simple model performing well.
\end{quotation}

\section{What else can go wrong?}

We can't let this opportunity pass to discuss the shortcoming of LOO methods without talking about the assumptions of LOO and how they can be violated. We've already mentioned that the assumption that the existing data is a good proxy for future observations can lead to problems. This is particularly true when data sets contain rare events.  But there is one other major assumption that LOO methods (and cross validation methods generally) make: the assumption of conditional exchangeability between observations (or groups of observations).

If the distribution of the observations, once conditioned on all of the model parameters, is not exchangeable LOO methods will fail.  This is because the fundamental idea of LOO is that $\frac{1}{n} \sum_{i=1}^n\log\left( \int\! p({y}_i| \theta) p(\theta | y_{-i})\,d\theta\right)$ is an unbiased approximation to the true expectation over unseen data $\mbox{E}_{\tilde{y}}\left( \log\left( \int\! p(\tilde{y}| \theta) p(\theta | _\text{\rm all})\,d\theta\right)\right).$ This only holds for models where observations are conditionally exchangeable.

This means that we need to be particularly careful when using cross-validation methods to assess models with temporal, spatial, or multilevel structure in the observations. Typically, we can alleviate this problem using a cross validation scheme that has been carefully designed to ensure that the blocks are (approximately) exchangeable (see, e.g., Roberts et al., 2017 and Bürkner et al, 2018).  Some evidence of correlation may be visible by plotting the difference in log predictive densities as a function of time or space (much in the same way we can check linear regression residuals for serial autocorrelation), as well as checking the individual models for uncorrelated residuals. Vivar and Ferreira (2009) present a visual check for cross-validation predictions.

To summarize our claim, we say that cross validation  methods will fail in all their tasks if the current data is not a good representation of all possible future data, or if the observations are not exchangeable conditional on the model parameters.  While stacking weights will consistently identify the correct predictive distribution, it will not necessarily choose the simplest model when multiple models asymptotically give the same predictive distribution. In the case of nested models, parsimony can be restored using projective prediction methods, however for non-nested models we don't know of a general solution.

\section{Can you do open science with $M$-closed tools?}

One of the great joys of writing a discussion is that we can pose a very difficult question that we have no real intention of answering. The question that is well worth pondering is the extent to which our chosen statistical tools influence how scientific decisions are made. And it's relevant in this context because of a key difference between model selection tools based on LOO and tools based on marginal likelihoods is what happens when none of the models could reasonably generate the data.

In this context, marginal likelihood-based model selection tools will, as the amount of data increases, choose the model that best represents the data, even if it doesn't represent the data particularly well. LOO-based methods, on the other hand, are quite comfortable expressing that they can not determine a single model that should be selected. To put it more bluntly, marginal likelihood will always confidently select the wrong model, while LOO is able to express that no one model is correct.

We leave it for each individual statistician to work out how the shortcomings of marginal likelihood-based model selection  balance with the shortcomings of cross-validation methods. There is no simple answer.

\section*{References}

\noindent

\bibitem Bengio, Y., and Grandvalet, Y. (2004). No unbiased estimator of the variance of $k$-fold cross-validation. {\em Journal of Machine Learning Research} {\bf 5}, 1089--1105.
  
\bibitem Bernardo, J. M., and Smith, A. F. M. (1994). {\em Bayesian Theory}.  New York:  Wiley.

\bibitem Breiman, L., Friedman, J., Olshen, R., and Stone, C. (1984). {\em Classification and Regression Trees}. London:  Chapman and Hall.

\bibitem Bürkner, P., Vehtari, A., and Gabry, J. (2018). Approximate leave-future-out cross-validation for time series models. \url{http://mc-stan.org/loo/articles/loo2-lfo.html}
  
\bibitem Dupuis, J. A., and Robert, C. P. (2003). Variable selection in qualitative models via an entropic explanatory power. {\em Journal of Statistical Planning and Inference} {\bf 111}, 77–-94.

\bibitem Gabry, J., Simpson, D. P., Vehtari, A., Betancourt, M., and Gelman, A. (2018).  Visualization in Bayesian workflow (with discussion). {\em Journal of the Royal Statistical Society A}.

\bibitem Gelman, A., Carlin, J. B., Stern, H. S., Dunson, D. B., Vehtari, A., and Rubin, D. B. (2013).
{\em Bayesian Data Analysis}, third edition.  London:  CRC Press.

\bibitem Gelman, A., Vehtari, A., and Hwang, J. (2014).  Understanding predictive information criteria for Bayesian models. {\em Statistics and Computing} {\bf 24}, 997--1016.
  
\bibitem Goutis, C., and Robert, C. P. (1998). Model choice in generalised linear models: A Bayesian approach via Kullback–Leibler projections. {\em Biometrika} {\bf 85}, 29-–37.

\bibitem Gronau, Q., and Wagenmakers, E. J. (2018).  Limitations of Bayesian leave-one-out cross-validation for model selection. {\em Computational Brain \& Behavior}.

\bibitem Neal, R. M. (1996).  {\em Bayesian Learning for Neural Networks}.  Lecture Notes in Statistics No.\ 118.  New York: Springer-Verlag.

\bibitem Piironen, J., and Vehtari, A. (2017). Comparison of Bayesian predictive methods for model selection. {\em Statistics and Computing} {\bf 27}, 711--735.

\bibitem Piironen, J, Paasiniemi, M., and Vehtari, A. (2018). Projective inference in high-dimensional problems: Prediction and feature selection. \url{https://arxiv.org/abs/1810.02406}

\bibitem Roberts, D. R., Bahn, V., Ciuti, S., Boyce, M. S., Elith, J., Guillera-Arroita, G., Hauenstein, S.,  Lahoz-Monfort, J. J., Schröder, B.,  Thuiller, W.,  Warton, D. I.,  Wintle, B. A., Hartig, F., and Dormann, C. F.  (2017). Cross-validation strategies for data with temporal, spatial, hierarchical, or phylogenetic structure. {\em Ecography} {\bf 40}, 913--929.
  
\bibitem San Martini, A., and Spezzaferri, F. (1984). A predictive model selection
criterion. {\em Journal of the Royal Statistical Society B}
{\bf 46}, 296--303.

\bibitem Simpson, D., Rue, H., Riebler, A., Martins, T. G., and Sørbye, S. H. (2017). Penalising model component complexity: A principled, practical approach to constructing priors. {\em Statistical Science} {\bf 32}, 1--28.

\bibitem Vehtari, A., Gelman, A., and Gabry, J. (2017). Practical Bayesian model evaluation using leave-one-out cross-validation and WAIC. {\em Statistics and Computing} {\bf 27}, 1413--1432. 

\bibitem Vehtari, A. (2018a).  Bayesian data analysis---beta blocker cross-validation demo.  \url{https://avehtari.github.io/modelselection/betablockers.html}

\bibitem Vehtari, A. (2018b).  Model assessment, selection and inference after selection.
\url{https://avehtari.github.io/modelselection/}

\bibitem Vehtari, A., and Gabry, J. (2018).  Using the loo package.  \url{https://cran.r-project.org/web/packages/loo/vignettes/loo2-example.html}

\bibitem Vehtari, A., and Ojanen, J. (2012). A survey of Bayesian predictive methods for model assessment, selection and comparison. {\em Statistics Surveys} {\bf 6}, 142--228.

\bibitem Vivar, J. C., and Ferreira, M. A. R. (2009). Spatiotemporal models for Gaussian areal data. {\em Journal of Computational and Graphical Statistics} {\bf 18}, 658--674.

\bibitem Wang, W., and Gelman, A. (2014).  Difficulty of selecting among multilevel models using predictive accuracy. {\em Statistics and Its Interface} {\bf 7}.

\bibitem Yao., Y., Vehtari, A., Simpson, D. P., and Gelman, A. (2018).  Using stacking to average Bayesian predictive distributions (with discussion). {\em Bayesian Analysis} {\bf 13}, 917--1003.

\end{document}